\documentclass[aps,groupedaddress,twocolumn,showpacs]{revtex4}
\usepackage{amssymb}

\begin{document}
\bibliographystyle{apsrev}


\title{CP violation and modular symmetries}



\author{Thomas Dent}
\affiliation{Michigan Center for Theoretical Physics,
      Randall Lab., Physics Department,\\
      University of Michigan, Ann Arbor, MI 48109 U.\ S.\ A.}


\date{\today}

\begin{abstract}
We reconsider the origin of CP violation in fundamental theory. Existing string models of spontaneous CP violation make ambiguous predictions, due to the arbitrariness of CP transformation and the apparent non-invariance of the results under duality. We find an unambiguous {\em modular CP invariance condition}, applicable to predictive models of spontaneous CP violation, which circumvents these problems; it strongly constrains CP violation by heterotic string moduli. The dilaton is also evaluated as a source of CP violation, but is likely experimentally excluded. We consider the prospects for explaining CP violation in strongly-coupled strings and brane worlds.
\end{abstract}

\maketitle

\section{Introduction: CP violation in and beyond the Standard Model} 
The origin of CP violation is an outstanding question in particle physics, which also has implications for atomic and nuclear physics and for cosmology, and must be addressed in any unified model. In this Letter we ask whether string theory can explain {\em why}\/ CP violation occurs and can predict observable CP-violation. The aim is to put limits on the space of string models, or conversely to motivate particular patterns of CP-violating low-energy parameters.

CP violation is allowed in the Standard Model (SM) via complex Yukawa couplings, which after spontaneous symmetry breaking result in a nonremovable complex phase $\delta$ in the CKM mixing matrix \cite{Kobayashi:1973fv}. Experimental results are currently consistent with this being the only source of CP violation, given a large (order 1) value of $\delta$; the small size of CP violating effects is explained by the small mixing angles of the CKM matrix, because CP violating observables always involve off-diagonal elements. The Yukawa couplings are not known and are free parameters in the SM, so the explanation of why CP violation occurs, and of the mixing angles, lies at higher energies. Experiments on B meson decays and electric dipole moments (EDM's) of fermions and atoms also have the ability to detect CP violating effects differing from the SM predictions, and the fact that the Universe has an excess of baryons over antibaryons indicates that the CKM phase is likely not the only source of CP violation, since no mechanism within the SM can create the observed asymmetry \cite{Rubakov:1996vz}. Thus it is of interest to search for the origin of CP violation in fundamental theory, both in order to explain the SM parameters and to look for effects beyond the SM. 

One should if possible use models that are calculable and predictive, meaning that the CP-violating effects are determined without depending on adjustable parameters put in by hand (or at least, that the number of testable predictions exceeds the number of input parameters) and also that the predictions are not subject to large unknown corrections. In contrast to the multiplicity of models proposed to explain the origin of quark masses and mixings (or ``flavour''), relatively few have addressed the origin of complex phases. The exceptions are models of spontaneous CP violation (SCPV) \cite{Lee:1973iz,Lee:1974jb,Weinberg:1976hu} in which the Lagrangian is CP invariant but the vacuum expectation values (v.e.v.'s) of some scalars are not, leading to effective CP-violating couplings at low energy. Gauge theory SCPV models require a symmetry-breaking potential and couplings of the scalars to the SM fields which are both inserted {\em ad hoc}, although they are generally constrained by other experimental data. In addition, in the absence of supersymmetry (SUSY) the scalar potential that stabilizes the CP-odd v.e.v.'s suffers large radiative corrections, reducing the predictivity of such models (see however \cite{Branco:1984tn}). In SUSY gauge theory the possibilities are extremely constrained and seem to rely on a particular choice of soft SUSY-breaking terms \cite{Branco:2000dq}. 

In contrast, compactified superstring theory provides, in principle, a class of models which meet our criteria. CP is a discrete gauge symmetry in string theory, so its violation is necessarily spontaneous and may depend on geometric parameters of the compact space $X$. In the low-energy field theory these parameters appear as {\em moduli}\/, scalar fields massless to all orders in perturbation theory which generically have imaginary, CP-odd components, since they live in chiral supermultiplets. After supersymmetry-breaking the scalar potential is lifted and the moduli v.e.v.'s are determined by the form of the effective potential, which is known for many string models. Yukawa couplings and soft SUSY-breaking terms are then calculable in terms of the topology of $X$ and the values of the moduli, so the couplings of CP-odd scalars to the SM fields (and superpartners) are explicitly determined in particular string models.

String models are particularly interesting in the context of supersymmetric extensions of the SM, which have a naturalness problem associated with CP violation. If the CP-violating phases introduced by the soft SUSY-breaking terms are of order 1, as would be expected by analogy with the CKM phase, then the EDM's of the neutron, electron and mercury atom are predicted to be outside the current experimental bounds by two or three orders of magnitude, even if flavour-changing effects are sufficiently suppressed: see \cite{Barger:2001nu} for a recent review. The ``supersymmetric CP problem'' seems to require the soft phases to be fine-tuned to be small, which is unnatural since no symmetry is restored when the phases vanish. Alternative proposals include approximate CP \cite{Eyal:1998bk,Dine:2001ne}, where all phases including $\delta$ are small and superpartner diagrams contribute significantly to the observed CP violation, large superpartner masses \cite{Kizukuri:1992nj,Cohen:1996vb} which suppress the loops contributing to EDM's, and cancellations between relatively large phases \cite{Ibrahim:1998gj,Brhlik:1998zn}. These proposals require particular patterns of soft terms or of CP-violating phases, which should have some motivation, rather than being an arbitrary choice in parameter space: string models can give such relations \cite{Brhlik:1999pw} since all the soft parameters are in principle determined by the moduli values.

Given the complexity of most (semi-)realistic string constructions, only toy models currently exist, based on heterotic string orbifolds \cite{Bailin:1997fh,Bailin:1998iz,Bailin:1998xx,Giedt:2000es}; they do however include the essential features of complex moduli stabilized after SUSY-breaking, which in turn determine the relevant low-energy couplings. However, even these simple models are somewhat difficult to interpret: the assignment of the Standard Model matter to string states is ambiguous, so the CP transformation is only defined up to a unitary matrix; thus it is nontrivial to decide whether CP violation actually occurs or not. The models also possess a target-space duality symmetry acting on the moduli and the matter fields, which usefully constrains the form of the scalar potential and couplings; however, the results for low-energy parameters are not manifestly duality-invariant, due to the nontrivial transformation of the observable fields. 


\subsection{An unambiguous condition for CP violation}
As a first step towards describing the CP phenomenology of string theory, we need a clear criterion of whether or not CP is broken by scalar field v.e.v.'s in the presence of such nontrivial duality symmetries. In this paper we use the freedom to use a unitary redefinition of matter fields to construct a condition on these v.e.v.'s for CP violation to occur. The {\em modular CP invariance condition}\/ allows the orbifold toy model results to be easily interpreted, and provides a strong constraint on the models. It generalises conditions for CP violation in gauge theory SCPV models \cite{Branco:1984tn} to any field theory model where the dynamics and couplings of the CP-odd scalars are constrained by some symmetry.

For a model of CP violation to be predictive, the relevant scalars should transform under some symmetry group, which also constrains the modulus-dependent couplings of the SM fields (we stretch the usual string theory definition of ``modulus'', as a scalar parameter with zero potential in perturbation theory, to include all scalars which may contribute to spontaneous CP violation). We take the SL$(2,\mathbb{Z})$ duality of some heterotic string compactifications \cite{Dijkgraaf:1987jt} as our main example, in common with existing models \footnote{The cancellation of all possible SL$(2,\mathbb{Z})$ anomalies in heterotic orbifolds has recently been shown by explicit string calculations \cite{Scrucca:2000ns}.}. Then CP is violated only by modulus values for which a CP transformation is not in the internal symmetry group, independently of the exact form of the modulus-matter couplings. For the SL$(2,\mathbb{Z})$ symmetry, the condition is surprisingly restrictive and rules out some proposals for CP violation phenomenology. The result should also be useful to evaluate other models of the origin of CP violation.

\subsection{CP in string theory and phenomenology}
CP is a discrete gauge symmetry \cite{Choi:1993xp,Dine:1992ya} in a class of higher-dimensional theories, including the SO$(32)$ and E$_8\otimes$E$_8$ string theories, and so is not explicitly broken, even by quantum gravity effects \cite{Krauss:1989zc}. Thus, CP violation is {\em spontaneous}\/, and in principle calculable. It is possible to break CP by compactifying on a Calabi-Yau manifold defined by complex parameters \cite{Strominger:1985it} but the dynamics of these parameters, corresponding to moduli in the $d=4$ effective theory, are difficult to calculate, and no explicit models of this type exist. Compactification on an orbifold preserving $\mathcal{N}=1$ SUSY respects CP symmetry \cite{Kobayashi:1995ks}, except for v.e.v.'s of CP-odd scalars in the effective field theory, which is well-known in this case \cite{Derendinger:1992hq}\footnote{It is conceivable that CP violation could result from a compactification with {\em discrete}\/ CP-odd parameters, or from scalars whose v.e.v.'s are determined near the string scale rather than surviving as flat directions. These possibilities deserve further investigation.}.

As potential sources of CP violation, as well as charged Higgses, the effective field theory for heterotic orbifolds contains the gauge singlet dilaton, $S$, and modulus, $T^\alpha$, scalars, whose axionic components Im$\,S$ and Im$\,T^\alpha$ correspond to the v.e.v.\ of the antisymmetric tensor in 4-d spacetime and in the compactified directions respectively. These components are odd under CP transformation, which reverses the orientation of spacetime and of the compactification manifold \cite{Choi:1993xp,Strominger:1985it}. CP violation might also arise through v.e.v.'s of scalars charged under a pseudo-anomalous U$(1)$ \cite{Giedt:2000es} or by twisted moduli associated to orbifold fixed points, although these have the drawback that the dynamics determining the CP-odd v.e.v.'s are unclear\footnote{{\em e.g.}\/\ if twisted fields receive v.e.v.'s, the compact space is no longer an orbifold, but is ``blown up'' into a Calabi-Yau, or may not even have a geometrical interpretation.}. 

Yukawa couplings in the supergravity effective theory are holomorphic functions of string moduli \cite{Lauer:1989ax}, so complex v.e.v.'s for the $T^\alpha$ may lead to CP violation by the CKM phase, as in the SM \cite{Bailin:1998xx}. Soft SUSY-breaking terms are also expected to break CP through complex v.e.v.'s of the auxiliary fields $F^S$ and $F^T$ \cite{Ibanez:1991qh,Bailin:1997fh}. The QCD vacuum angle is proportional to Im$\,\langle S\rangle$ at tree level, so if non-zero it must be cancelled by other contributions or by an axion \cite{Ibanez:1991qh,Choi:1997an}, to satisfy the bound $\bar{\theta}_{\rm QCD}\leq 10^{-10}$ arising from measurements of the EDM of the neutron.
Complex Yukawa couplings and soft terms have been shown to arise in orbifold models \cite{Bailin:1997fh,Bailin:1998iz,Bailin:1998xx} when the moduli are stabilized after supersymmetry-breaking by gaugino condensation, in the presence of $T$-dependent threshold corrections. These are best regarded as toy models, since they have an unrealistic gauge group and matter spectrum; the assignment of the SM fields will depend on further stages of symmetry-breaking, which may however introduce further sources of CP violation \cite{Bailin:1998yt}. Stabilization of the dilaton and moduli (including twisted moduli) can also be achieved in Type I models \cite{Abel:2000tf}. 

Even toy models are significantly constrained by the target-space duality acting on the $T$ moduli and by the possibility that complex phases may be removed by redefining the matter fields \cite{Bernabeu:1986fc}. For example, Giedt \cite{Giedt:2000es} showed that CP was in fact conserved for a particular set of complex Yukawa couplings \cite{Bailin:1998xx}, assuming particular assigments of the SM fields to string states. The modular CP invariance condition constructed here explains such results without having laboriously to consider all possible field redefinitions. For the symmetry group SL$(2,\mathbb{Z})$ we confirm the conjecture \cite{BKL_unpub} that values of $T$ on the boundary of the fundamental domain $\mathcal{F}$ are CP-conserving, from which the result of \cite{Giedt:2000es} follows as a special case. We also address the problem of the apparent {\em modular non-invariance}\/ of modulus-dependent coupling constants, which occurs when the matter fields transform non-trivially, and consider the possibility in the heterotic string that CP may be conserved by $T$-dependent couplings but violated by Im$\,\langle S\rangle$. Finally we look at other classes of semi-realistic string models and, briefly, more general models involving extra dimensions.

\section{Modular symmetry and the low-energy effective theory} 
The CP invariance condition will be constructed with as few assumptions as possible, then applied to the example of SL$(2,\mathbb{Z})$ invariant compactifications. For the purpose of discussion we divide up the fields in the effective action $\Gamma$ into ``moduli'' $\hat{T}$, classical fields which determine the value of low-energy couplings and which are to receive v.e.v.'s, and observable fields $\Phi$, whose excitations can be produced and detected. Thus a Higgs doublet may be written as $\phi = (H^+, v+H^0)$ with $v=\langle\phi\rangle$, such that $v$ belongs to $\hat{T}$ and $H^+$, $H^0$ to $\Phi$. Then, if the effective theory is invariant under a symmetry $\mathcal{M}$, we have
\begin{equation}
\Gamma(\mathcal{M}(\hat{T}),\mathcal{M}(\Phi)) = \Gamma(\hat{T},\Phi)\ \forall \mathcal{M} \label{eq:modinv}
\end{equation}
where $\mathcal{M}$ acts on  $\mathcal{M}(\hat{T})$ and $\mathcal{M}(\Phi)$ represent the action of $\mathcal{M}$ on moduli and observable fields respectively. For the example of SL$(2,\mathbb{Z})$ duality, we have for the overall volume modulus $T$
\begin{equation}
\mathcal{M}:\ T\mapsto (\alpha T-i\beta)(i\gamma T+\delta)^{-1} \label{eq:Ttr_SL2Z},
\end{equation}
and for chiral matter superfields $U$ and $A$ (untwisted and twisted respectively)
\begin{equation} 
\mathcal{M}:\ U\mapsto (i\gamma T+\delta)^{-1}U_i,\ A_a\mapsto M_{ab} (i\gamma T+\delta)^{n_a} A_b \label{eq:matter_SL2Z}
\end{equation}
where the group element $\mathcal{M}$ is specified by integers $\alpha$, $\beta$, $\gamma$ and $\delta$ with $\alpha\delta-\beta\gamma=1$, the constant $n_a$ is the modular weight of the field $A_a$ and the unitary matrix $\mathbf{M}$ depends on $\mathcal{M}$ but not on $T$. The transformations (\ref{eq:Ttr_SL2Z}), (\ref{eq:matter_SL2Z}) are easily generalised to the case of two or more independent moduli. At one loop in string theory the dilaton $S$ is also shifted under SL$(2,\mathbb{Z})$ \cite{Derendinger:1992hq} (see eq.\ \ref{eq:Sshift}). For $\Gamma$ to be invariant, modulus-dependent couplings in the low-energy theory must transform, so duality invariance is spontaneously broken when $T$ receives a v.e.v.. The gauge charges of matter fields are unchanged under the duality and, although the action on the chiral superfields (\ref{eq:matter_SL2Z}) is nonunitary, the group acts unitarily on canonically-normalised component fields \cite{Ferrara:1989qb}.

If the symmetry group $\{\mathcal{M}\}$ is non-Abelian, individual terms in the action may not be invariant, and modulus-dependent couplings may mix with one another under modular transformations: {\em e.g.\/}\ in orbifold twisted sectors, for which the $M_{ab}$ in (\ref{eq:matter_SL2Z}) are nondiagonal \cite{Lauer:1989ax}. Results for CP-violating couplings then become ambiguous, since they appear to change under modular transformation (which should leave physics invariant). We anticipate that this problem may be solved by a modulus-dependent change of basis of observable fields ({\em e.g.\/}\ a transformation to the mass eigenstate basis), such that the redefined fields are invariant under the combined action of (\ref{eq:Ttr_SL2Z}) and (\ref{eq:matter_SL2Z}). Then the couplings of the redefined fields are {\em invariant}\/ functions of the moduli, even though modular symmetry is spontaneously broken. Note that nondegenerate mass eigenstates may not be mixed by $\mathcal{M}$ (except in the pathological case where the states are {\em permuted}\/, and all their modulus-dependent physical properties transform accordingly). 

\section{General CP transformations and the CP invariance condition}
The other ingredient in our construction is the concept of general CP (GCP) transformations \cite{Bernabeu:1986fc,Botella:1995cs}. CP symmetry can strictly be defined only for a CP-conserving action, so the potentially CP-violating interactions are to be treated as a perturbation to the CP-conserving action $\Gamma_g$, which will in general include all kinetic terms and minimal gauge couplings. CP is then only defined up to an element of the internal symmetry group $\{G\}$ of $\Gamma_g$, which may include constant gauge transformations, global symmetries and duality transformations. We write the action of a particular GCP transformation $\mathcal{GCP}[G]$ on the observable fields as $\mathcal{GCP}[G](\Phi) = \left(G(\Phi)\right)^{\rm CP}$, where the standard CP transformation is
\begin{eqnarray} &\psi_L \mapsto (\psi_L)^{\rm CP}=i\sigma^2\psi_L^*,\ \psi_R \mapsto (\psi_R)^{\rm CP}=-i\sigma^2\psi_R^*,& \nonumber \\ &\phi \mapsto (\phi)^{\rm CP}=\phi^*,\ V_\mu\mapsto (V_\mu)^{\rm CP}=-V^\mu.& \label{eq:nCP}
\end{eqnarray}
for Weyl spinors $\psi_L$, $\psi_R$, complex scalars $\phi$ and vectors $V^\mu$. Then CP is conserved if and only if there is at least one element $G^x$ such that $\mathcal{GCP}[G^x]$ leaves the effective action unchanged:
\begin{equation}
\mbox{CP is conserved}\Leftrightarrow\exists\,G^x: \Gamma(\hat{T},\mathcal{GCP}[G^x](\Phi)) = \Gamma(\hat{T},\Phi). \label{eq:CPcons}
\end{equation}
This is equivalent to finding some change of basis of observable fields that puts the action into a form invariant under the standard CP transformations (\ref{eq:nCP}).

Since CP cannot be explicitly violated, the action is invariant under a CP transformation acting on both the values of the moduli and on observable fields. In general, the CP transformation of moduli $\hat{T} \mapsto \hat{T}^{\rm CP}$ must be consistent with CP being a dicrete gauge symmetry. Then we have 
\begin{equation}
\Gamma(\hat{T}^{\rm CP}, \mathcal{GCP}[G^a](\Phi)) = \Gamma(\hat{T},\Phi) \label{eq:CPspont}
\end{equation}
for at least one $G^a$ ({\em n.b.\/}\ $G^a$ must satisfy $G^a(G^a)^{\rm CP}=\mathbf{1}$, where $(G^a)^{\rm CP}(\Phi^{\rm CP})\equiv (G^a(\Phi))^{\rm CP}$). In the heterotic string we have $S(T) \mapsto S^*(T^*)$ under CP, since the imaginary parts are CP-odd. 

If the group element $G$ is anomalous, the above statements for the effective action $\Gamma$ appear not to be strictly valid, since $G$ is not a quantum symmetry of the action $\Gamma_g$. Even if the perturbative couplings are GCP invariant, a term in $\Gamma$ proportional to Tr$\,F\tilde{F}$ for the SU$(3)_{\rm C}$ gauge group is generated by the action of $G$, so the effective $\theta_{\rm QCD}$ angle will be shifted, in addition to changing sign under CP. But this contradicts our belief that GCP transformations should be physically equivalent to the standard CP transformation (\ref{eq:nCP}). However, if we assume that the strong CP problem is solved by the presence of extra unobservable fields, then $G$ may be altered to include, for example, a shift in the axion, or chiral rotations of heavy fermions in a Nelson-Barr-like model, to cancel the shift in $\theta$.

Now consider the GCP transformation $\mathcal{GCP}[G^{a}\mathcal{M}^c]$ acting on $\Gamma(\hat{T},\Phi)$, where the group element $\mathcal{M}^c$ is chosen so that the {\em modular CP invariance condition}\/ 
\begin{equation}
	\hat{T}^{\rm CP} = \mathcal{M}^{c}(\hat{T}) \label{eq:modcond}
\end{equation}  
holds, for the values $\{\hat{T}^c\}$ of some subset $\hat{T}^c$ of the moduli. For these values of the moduli $\hat{T}^c$, a CP transformation is identical to a modular symmetry, under which $\hat{T}^c$-dependent physics should be invariant: CP {\em cannot be violated}\/ by couplings depending only on the $\hat{T}^c$. More precisely, we have
\begin{equation}
	\Gamma(\hat{T},\Phi)\stackrel{\mathcal{GCP}[G^{a}\mathcal{M}^c]}{\longmapsto} \Gamma^{\rm GCP} \equiv \Gamma(\hat{T},(G^{a})^{\rm CP}(\mathcal{M}^c(\Phi))^{\rm CP}). \label{eq:GCPGaMc}
\end{equation}
In order to define this transformation we require the action of $\mathcal{M}^c$ on the observable fields to be an element of $\{G\}$. But this action leaves gauge charges invariant and is unitary for properly-normalised fields: so (\ref{eq:GCPGaMc}) is a valid (G)CP transformation.

Then, using the statement of spontaneous CP violation (\ref{eq:CPspont}) we have
\begin{eqnarray}
	\Gamma^{\rm GCP} &=& \Gamma(\hat{T}^{\rm CP},\mathcal{M}^{c}(\Phi)) \\
	&=& \Gamma(\hat{T}^c,(\mathcal{M}^c)^{-1}(\hat{T}^{v\rm CP}),\Phi) \label{eq:GammaGCP3}
\end{eqnarray} 
where the last equality follows from modular invariance under the group element $(\mathcal{M}^{c})^{-1}$, and $\hat{T}^v$ are the moduli for which our condition (\ref{eq:modcond}) does {\em not}\/ hold. Then under this CP transformation, {\em only $\hat{T}^v$-dependent quantities change}\/, as shown in (\ref{eq:GammaGCP3}). This is our main result\footnote{We assume that $\bar{\theta}_{\rm QCD}$ vanishes throughout.}.

The CP invariance condition (\ref{eq:modcond}) is applicable to any symmetry group acting on CP-odd scalars. For example, in field theory SCPV models one may impose $\mathbb{Z}_N$ symmetries acting on the Higgses, and CP is unbroken for vacua which are connected to their CP conjugates by a $\mathbb{Z}_N$ transformation. Our method extends such well-known examples to highly nontrivial symmetries of the type found in string theory, and does not require complete specification of the model.

\section{Applications} 
Now consider the heterotic orbifold models with duality group SL$(2,\mathbb{Z})$ where the $T$ moduli are the candidate sources of CP violation. By using for $\mathcal{M}^c$ the group generators $T\mapsto 1/T$, $T\mapsto T+i$, we see that CP is conserved for $T$ on the unit circle and on the lines Im$\,T=\pm 1/2$; that is, precisely {\em on the boundary of the fundamental domain}\/ $\mathcal{F}$. Thus the result \cite{Giedt:2000es} that the complex phases of Yukawa couplings for the v.e.v.'s $T=e^{\pm i\pi/6}$ in a particular string model do not result in CP violation, follows immediately. So, if the v.e.v.\ of $T$ is the source of CP violation then it must lie inside the fundamental domain. This is difficult to achieve for stabilization mechanisms respecting modular invariance \cite{Cvetic:1991qm} but possible using gaugino condensation with universal threshold corrections \cite{Bailin:2000ra,Nilles:1997vk}. Thus, considering CP violation can give clues about other aspects of the fundamental theory.

The solution of the supersymmetric CP problem by complex $T$ values lying on the boundary of $\mathcal{F}$ \cite{Ibanez:1991qh}, leading to soft terms that vanish or conserve CP, does not give a consistent picture of CP violation, since the Yukawa couplings would then also conserve CP. Conversely, the proposal that soft SUSY-breaking terms are the only source of CP violation \cite{Abel:1997eb,Brhlik:1999hs} is not motivated either in this scenario (unless the KM phase vanishes by chance for $\langle T\rangle$ at isolated points within $\mathcal{F}$). For generic modulus v.e.v.'s inside $\mathcal{F}$, we expect CP violation in all possible sets of couplings, favouring the proposals of approximate CP or relatively large soft phases which evade the EDM bounds by cancellations or nontrivial flavour structure \cite{Ibrahim:1998gj,Brhlik:1998zn}. Approximate CP would result from modulus v.e.v.'s close to the CP-conserving boundary of $\mathcal{F}$. It is worth remarking that approximate CP is a generic possibility in string models that have light moduli, given that the source of CP violation is the moduli v.e.v.'s.

We can also determine whether the dilaton v.e.v.\ can lead to an acceptable pattern of CP violation in the heterotic string. The shift
\begin{equation}
	\mathcal{M}:\ S\mapsto S-\frac{3\delta_{\rm GS}}{(8\pi^2)} \ln(i\gamma T+\delta) \label{eq:Sshift}
\end{equation}
under modular transformations is inconsistent with the CP invariance condition (\ref{eq:modcond}) unless Im$\,S$ takes particular isolated values, which would be unlikely given a particular stabilization mechanism. So $S$ may be in the subset $\hat{T}^v$ of CP-violating ``moduli'', unless there is an axionic symmetry which removes the dependence of low-energy physics on Im$\,S$. Such a symmetry is desirable for the dependence of $\theta_{\rm QCD}$, but it is likely to be strongly broken \cite{Choi:1997an} by the dependence of SUSY-breaking on Im$\,S$. Then, for $\langle T\rangle$ on the boundary of $\mathcal{F}$ the CP invariance condition for $S$ is
\[ S^*=S-3\Delta_{\rm GS}\ln(i\gamma T+\delta) \]
where $\mathcal{M}^c(\alpha,\beta,\gamma,\delta)$ takes $T\mapsto T^*$. If Im$\,\langle S\rangle$ does not satisfy this then CP would be violated {\em by the $S$-dependence of the low-energy couplings only}. The Yukawa couplings, which in string theory do not depend on Im$\,S$, would conserve CP, while soft SUSY-breaking terms would break it via a complex $F^S$. Unfortunately, the dilaton-dependent CP violation that arises in the heterotic string is of the ``universal'' type, and appears to be inconsistent with generating sufficiently large CP violation in the kaon system while respecting the experimental bounds from fermion EDM's \cite{Brhlik:1999hs,Ibrahim:1998gj}.


Finally we examine the prospects for finding the origin of CP violation in fundamental theory beyond the heterotic string, and ask to what extent our method is applicable. The obvious extension is to heterotic M-theory, defined as the strong coupling limit of the heterotic string \cite{Horava:1996qa}. For vacua which can be continuously connected to the perturbative heterotic string by varying the size of the 11th dimension, CP should survive as a discrete gauge symmetry and target-space modular invariance should also hold (see \cite{Nilles:1997vk}), so the result would still be applicable, with the appropriate reinterpretation of the dilaton and $T$ moduli . Calculations of SUSY-breaking terms in the M-theory limit of large $S$ and $T$ show exponentially vanishing imaginary parts \cite{Bailin:2000ra}, so even if $\langle T\rangle$ lies inside $\mathcal{F}$ the phases in soft terms may be negligible. However, more work is needed to find the behaviour of the Yukawa couplings in this region. For non-standard embeddings including fivebranes, the status of modular invariance and CP violation remains to be determined, although recent work on the effective action \cite{Derendinger:2000gy} suggests that the {\em r\^ole}\/ of the fivebrane moduli may be similar to that of $S$ and $T$, as complex scalars potentially contributing to spontaneous CP violation.

CP is also a discrete gauge symmetry in the perturbative Type II string \cite{Dine:1992ya}, but may not survive as such in the presence of general D-brane configurations, since the gauge group is altered. Then explicit CP violation by essentially stringy or nonperturbative effects might occur, so an axion would be needed to solve the strong CP problem and the calculability of CP-violating phases might be affected. Taking the conservative position that nonperturbative effects do not contribute significantly, then the picture of spontaneous CP violation through string background fields survives and it would still be possible to build effective field theory models. The status of duality symmetries is slightly different from heterotic string: type I/type IIB models are known to have T-duality symmetries that exchange the dilaton and different moduli (see {\em e.g.\/}\ \cite{Ibanez:1998rf}) in the effective field theory. In addition the effective supergravity has SL$(2,\mathbb{R})$ symmetries acting on the compactification moduli, with sigma-model anomalies cancelled by shifts in the twisted moduli rather than the dilaton \cite{Ibanez:1999pw,Scrucca:2000za}. If the sigma-model symmetry (or a subgroup thereof) persists as a symmetry of the full theory, then our result can be carried over in a rather simple way; however, comparison with one-loop threshold corrections \cite{Lalak:1999ex} indicates that the situation is likely to be more complicated. The origin of complex Yukawa couplings is also a puzzle in these models, since there is apparently no holomorphic modulus-dependence in the perturbative superpotential.

More general ``brane world'' models, in which matter fields are localized in extra-dimensional space, also allow for CP violation \cite{Huang:2001np,Sakamura:2000ik,Chang:2001yn}, but the models use more or less arbitrary assumptions about the sources of CP violation, which lessen their predictivity. To put them on a more definite footing, one should describe the dynamics of the extra dimensions and of the localized ``branes''. In essence it should be shown that the parameters which are supposed to break CP actually take up the required values, within a self-contained model. Then one should use a condition analogous to (\ref{eq:modcond}) in order to evaluate the models.


%
%

\begin{acknowledgments}
The author thanks David Bailin for the original motivation of this work and for many discussions of modular invariance, and Gordy Kane, Dan Chung and Lisa Everett for helpful comments. Research supported in part by DOE Grant DE-FG02-95ER40899 Task G.
\end{acknowledgments}


\end{document}